\documentclass[10pt]{article}

\usepackage{graphicx}
\usepackage{amsmath}
\usepackage{verbatim}

\title{An early warning method for crush}
\author{
Peter J. Harding$^{1}$,  Steve M. V. Gwynne$^{2}$ and Martyn Amos$^{1,*}$ \\ \\
\small $^{1}$ Dept. of Computing and Mathematics, Manchester Metropolitan University, UK. \\
\small $^{2}$ Hughes Associates, Inc., USA. \\
\small $^{*}$ Corresponding author: M.Amos@mmu.ac.uk
}

\begin{document}

\date{}

\maketitle

\begin{abstract}
Fatal crush conditions occur in crowds with tragic frequency. Event organizers and architects are often criticised for failing to consider the causes and implications of crush, but the reality is that the prediction and mitigation of such conditions offers a significant technical challenge. Full treatment of physical force within crowd simulations is precise but computationally expensive; the more common method of human interpretation of results is computationally ``cheap" but subjective and time-consuming. In this paper we propose an alternative method for the analysis of crowd behaviour, which uses information theory to measure crowd disorder. We show how this technique may be easily incorporated into an existing simulation framework, and validate it against an historical event. Our results show that this method offers an effective and efficient route towards automatic detection of crush.
\end{abstract}

\section{Introduction}
\label{intro}
Overloading pedestrian routes can quickly lead to the development of {\it crush conditions}, as observed in the Hillsborough  \cite{taylor1989hsd}, Station nightclub  \cite{station} and Saudi Arabian Hajj  \cite{helbing2007dcd} incidents, as well as the recent Love Parade tragedy in Germany. A more sophisticated understanding of how crush conditions form is therefore critical for the design of tall buildings and other highly-populated, contained regions (such as ships, nightclubs and stadia), as well as for the planning of events and formulation of incident management procedures. A first step towards this deeper understanding is a method for {\it detecting} the early-stage formation of crush, which is the problem we address here.

The study of crowd evacuation/control scenarios has taken on additional significance in the light of events such as 9/11 . Many tall buildings (such as the World Trade Center towers) were designed alongside the assumption that any necessary evacuation could and would be conducted in a {\it phased} manner (e.g. floor-by-floor). One significant factor in building design is the capacity of exit routes (such as corridors and stairwells). Capacities are calculated based on projections of {\it controlled} population movement in phased evacuations. If the phased evacuation assumption breaks down (if, for example, occupants of a specific floor refuse to wait their ``turn" for fear of catastrophic building failure) then this will have severe implications for overall safety, as exit routes become overloaded. 

Computer-based simulation studies are often used to analyse the movement of individuals in various scenarios. Such work encompasses the study of historical events  \cite{helbing2007dcd}, the examination of evacuation procedures  \cite{gwynne2003a}, and the design of aircraft  \cite{galea03}. Existing simulation frameworks include EXODUS  \cite{owen1996exodus}, PEDFLOW  \cite{kukla2001pda} and EVACNET  \cite{kisko1985ecp} (see  \cite{kuligowski2005rem} for an extensive review), and these offer a range of ``real world" features, including exit blockage/obstacles, occupant impatience and route choice  \cite{gwynne1999a}. However, the phenomenon of {\it crush} is one that has received relatively little attention so far from the designers of evacuation simulations. Many simulations do not {\it explicitly} consider the effects of crush, and those that {\it do} factor in crush employ computationally expensive physical force calculations.

The two major problems we address are as follows: firstly, the consideration of crush within existing simulation frameworks requires the use of computationally intensive Newtonian force calculations. These can drastically slow down simulations, restricting their applicability in the rapid prototyping of building designs and crowd control procedures. The second problem is that the monitoring of crush within real crowds is rudimentary, at best, and relies largely on personal observation and interpretation of crowd patterns. This method of crush detection is inherently problematic.

We therefore seek a method for the detection of crush conditions that is relatively ``cheap" in terms of computational effort, and which may be easily integrated into existing software for crowd monitoring. Such a method will have a significant impact on both simulation-based evacuation studies and real-time analysis of video images (facilitating, for example, the development of automated crush alarms based on CCTV images). In this paper we give a description of our proposed method, which is based on the notion of {\it phase transitions} in a system of interacting particles. We show how our method may be easily integrated into an existing simulation framework, and test it using details of an historical event. Our results show that mutual information provides an excellent ``early warning" indicator of the emergence of crush conditions.

The rest of this paper is organised as follows: in Section ~\ref{previous} we first define the notion of ``crush conditions" , and examine how crush has been handled by previous simulation studies. This motivates the search for a new crush detection method, and we show in Section ~\ref{ourapproach} how the concepts of {\it phase transition} and {\it mutual information} might usefully be applied to the detection (and prediction) of crush. We describe the results of experimental investigations in Section ~\ref{results}, and conclude in Section ~\ref{conc} with a discussion of open questions.
	
\section{The problem of crush}
\label{previous}

We first consider the notion of crush conditions. As Fruin observes  \cite{fruin93}, 
overcrowding can often lead to injuries and/or fatalities; these {\it may} be caused by trampling or falls, but here we are concerned with the particularly common phenomenon of {\it compressive asphyxia} (also known as {\it chest compression}), which, Fruin argues, is responsible for ``virtually all crowd deaths"  \cite{fruin93}. This occurs when the torso is compressed by external forces, preventing expansion of the lungs and thus interfering with normal breathing. Difficulty in breathing due to intense pressure levels can often be exacerbated by anxiety and heat, quickly leading to significant physiological problems.

Fatal levels of force can emerge within a crowd as a result of pushing, leaning or (less commonly) vertical stacking of bodies. Images of steel barriers bent out of shape (for example, in the aftermath of the Hillsborough disaster \cite{taylor1989hsd}) graphically illustrate the extent to which force levels can grow. Fruin reports the results of several studies (either after-the-event forensic tests, or controlled experiments) which suggest that forces exceeding around 1500N could prove fatal  \cite{fruin93}. It is therefore an important factor to be considered in simulation studies aimed at improving structural designs or evacuation/control procedures, along with other aspects such as panic or physical obstacles.
	
Crush detection methods used to date in simulation studies may be classified into two generic groups; \textit{explicit methods} and \textit{implicit methods} cite{harding2008a}.  The \textit{implicit methodology} is the traditional approach, and is still highly popular, being the preferred technique in a large number of simulation models  (see \cite{kuligowski2005a} for an extensive review).  It relies on the {\it expert analysis} of factors such as population density and environmental considerations, yielding a {\it human interpretation} of the output of the simulation to help determine whether or not crush might have occurred. Although subjective, this method is still popular, because it does not require the use of computationally expensive force calculations, relying instead on human expertise and intuition. 

The \textit{explicit modelling} of crush conditions incorporates an assessment of crush into the model itself, and therefore requires less human analysis than the implicit approach.  Usually based on the calculation of Newtonian force values, and operating in 2-dimensional space, explicit methodologies are used to detect the presence of crush conditions in a much more objective fashion.  By simulating the physical force exerted by each individual, they calculate the precise amount of force present within a crowd.

Whilst the explicit methodologies offer a measure of the forces acting within a crowd, the calculations needed to assess levels of force require much more computer processing power than an implicit method. Experiments show that the computation time required by a model that {\it explicitly} quantifies force can be up to 100 times greater than that required by an {\it implicit} model \cite{quinn2003a}.

Given the nature of the current trade-off between precision and computational cost, we therefore seek a relatively ``cheap" method (in terms of run time) that will allow us to {\it automatically} signal the onset of crush conditions within an evacuation.  This will bridge the gap between the two current extremes, allowing architects and policy-makers to quickly and easily incorporate crush into their simulation scenarios. In the next Section, we explain how this may be achieved using Mutual Information.

\section{Mutual information for the detection of crush}
\label{ourapproach}

While studying video footage of the 2006 Saudi Arabian Hajj disaster, in which over 340 pilgrims died as the result of a stampede, Helbing {\it et al.} noticed distinct {\it transitions} in the flow of pedestrians around the time of the significant incident. They observed ``a sudden transition from laminar to ... unstable flows"  \cite{helbing2007dcd}; that is, a sudden ``flip" from {\it smooth} to {\it irregular} flows of human movement. Such transitions are, we believe, key to the early detection of crush, and we now describe our proposed methodology for their detection.
	
Our proposal is that the onset of crush can be detected via the analysis of crowd behaviour.  More specifically, by identifying {\it sustained periods of disorder}, we may identify the possible onset of crush.  By treating analysing this change in observable behaviour using information theory, we qualify the onset of crush conditions without ever {\it explicitly} calculating the amount of force present in the simulation.
	
Within a simulation, the two distinct states of a crowd are characterised by the behaviour of  individuals.  Under ``normal" conditions, crowd flow is highly ordered, with the orientation and speed of a specific individual being similar to that of those in their immediate locality.  The onset of more turbulent flow sees individuals exhibit a marked change in behaviour, as they change speed and alter course in order to avoid others.  We therefore wish to identify these distinct states, and we achieve this by applying statistical analysis techniques to the movement of individuals within crowds. 

\subsection{Mutual Information}
\label{mi}

Mutual Information (MI) is a probabilistic method for quantifying the interdependence of two variables.  It has previously been employed as an analytical technique in many areas  \cite{fraser1986a,jeong2001a,thevanez2000a}. More recently, it has been shown that MI may be used to identify a kinetic phase transition in a complex, dynamical system of interacting particles \cite{vicsek1995a}.  It is therefore possible to reliably identify the point at which certain particle-based systems move away from a disordered state and begin to exhibit some degree of order \cite{wicks2006a}, and vice versa (this is the phase transition).

In the general case, the Mutual Information of two \textit{discrete} time-series variables, A and B, is defined as:

	\begin{equation}
		I(A,B)  =  \sum_{i,j}p(a_i,b_j)\log_{n}\frac{p(a_i,b_j)}{p(a_i)p(b_j)}
	\label{eq:mi}
	\end{equation}

where $p(a_i)$, $p(b_j)$, and $p(a_i,b_j)$ are the individual probability and joint probability distributions of $A$ and $B$. In general terms, MI quantifies the interdependence of two variables; therefore if A and B are entirely {\it independent}, then $I(A,B) = 0$, but in \textit{all} all other cases $I(A,B) > 0$. In the next Section, we show how MI may be integrated with an existing simulation framework, to provide an entirely new metric for the analysis of pedestrian evacuation.

\section{Experimental investigations}
\label{results}

In this Section we describe the results of experiments to investigate the applicability of MI as a plausible tool for crush detection. In order to ensure its broad applicability, we first show how MI may be easily integrated into an existing, industry-standard simulation framework. We then validate the technique, by using it to analyse an historical event. By demonstrating that the MI technique correctly detects known incidences of crush within this scenario, we provide support for its adoption as a standard tool.

The base simulation environment used is the Fire Dynamics Simulator (FDS) \cite{ryder2004consequence}, a fluid dynamics-based model of fire and smoke flow. The {FDS+Evac} module \cite{korhonen2007fds+} is an evacuation simulation extension for FDS, and is based on the well-known {\it social forces model} \cite{helbing1995a,helbing2000a} ({SFM}) of pedestrian movement.

The evacuation module for FDS incorporates the calculation of physical forces, negating the need for additional functionality in this respect.  The MI analysis was integrated into the FDS environment as a set of natively coded (FORTRAN 90) libraries.  As the technique is entirely passive, i.e. it will not affect the results of the evacuation, there were no concerns regarding the effect this could have on the {\it behaviour} of the simulations (although there is clearly a small overhead incurred by the MI calculations).

The MI of the system is calculated at every \textit{simulation} time step, and the results averaged over 100 time steps before being recorded. This equates to one MI reading per second of \textit{real-life} evacuation time, which gives sufficient granularity. We record the average physical force within a simulation in the same way.

\subsection{Experimental validation}

In order to validate the technique, we choose a well-documented incident that illustrates the significant hazards that an emergency evacuation may present.  In 2003, the Station Nightclub (Rhode Island, USA) was the scene of one the worst nightclub fires in recent history, when a pyrotechnic device, used by the rock band Great White, ignited sound insulation foam in the walls and ceiling of the venue.  According to the official report into the incident \cite{station}, a crush formed at the main escape route within 90 seconds of the start of the fire, trapping patrons inside the club as it filled with smoke.  Estimates of the nightclub occupancy vary between 440 and 460; a total of 96 people died during the incident.

We select this particular event on the basis of (a) the existence of a significant amount of professional film footage taken inside the nightclub during the incident\footnote{Ironically, the film crew was present to record a documentary on nightclub safety, after a fatal incident elsewhere four days previously.}, (b) availability of supporting witness evidence and other associated documentation, and (c) results from substantial simulation tests using FDS as part of the subsequent (extensively documented) formal investigation. We therefore have information on the initial distribution of individuals at the {\it beginning} of the incident, visual evidence of crush {\it during} the incident, and the {\it final locations} of each of the victims, as well as a set of validated simulations with which to compare our own results.

\begin{figure}
		\centering
			\includegraphics[width=0.7\linewidth]{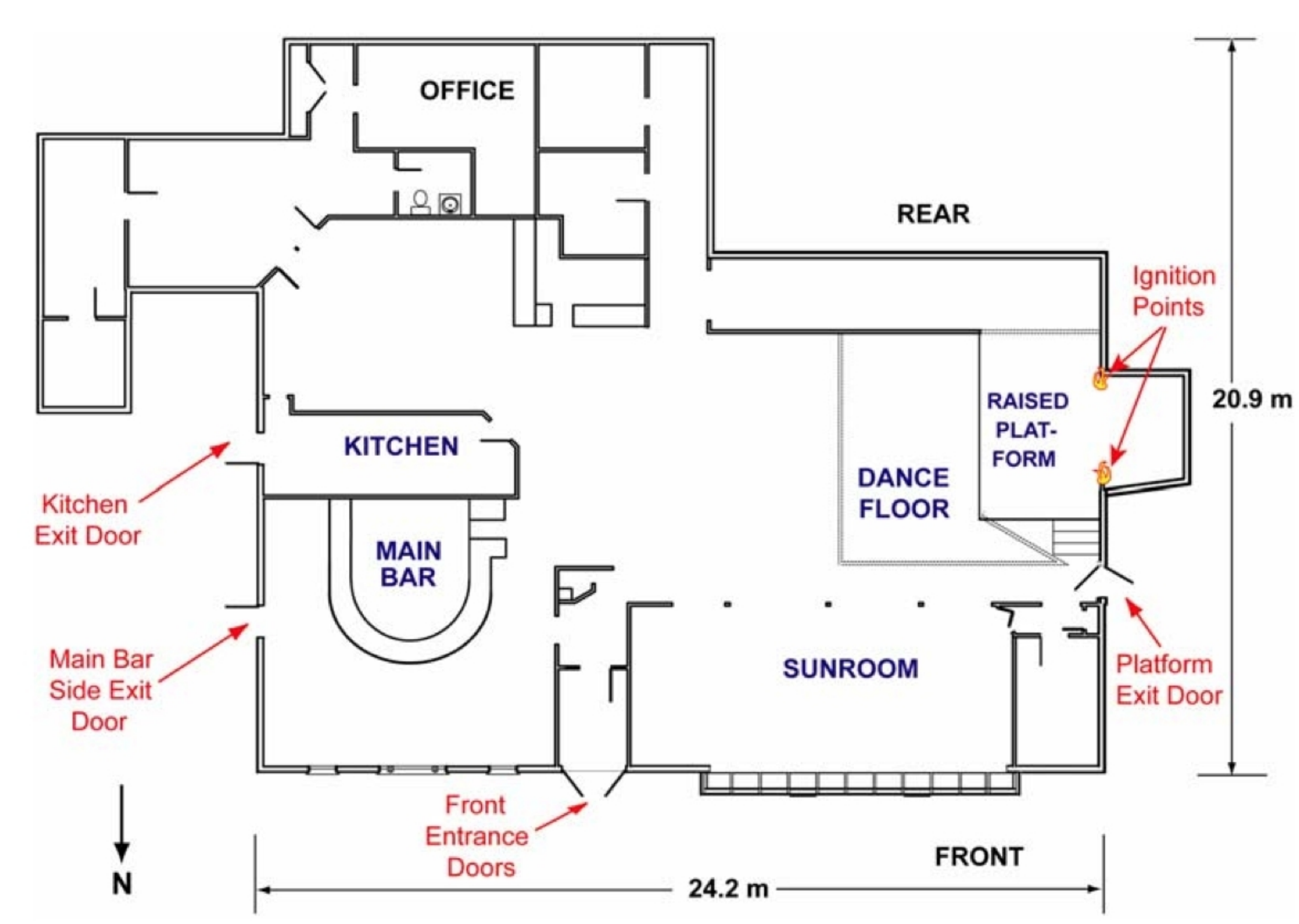}
			\includegraphics[width=0.7\linewidth]{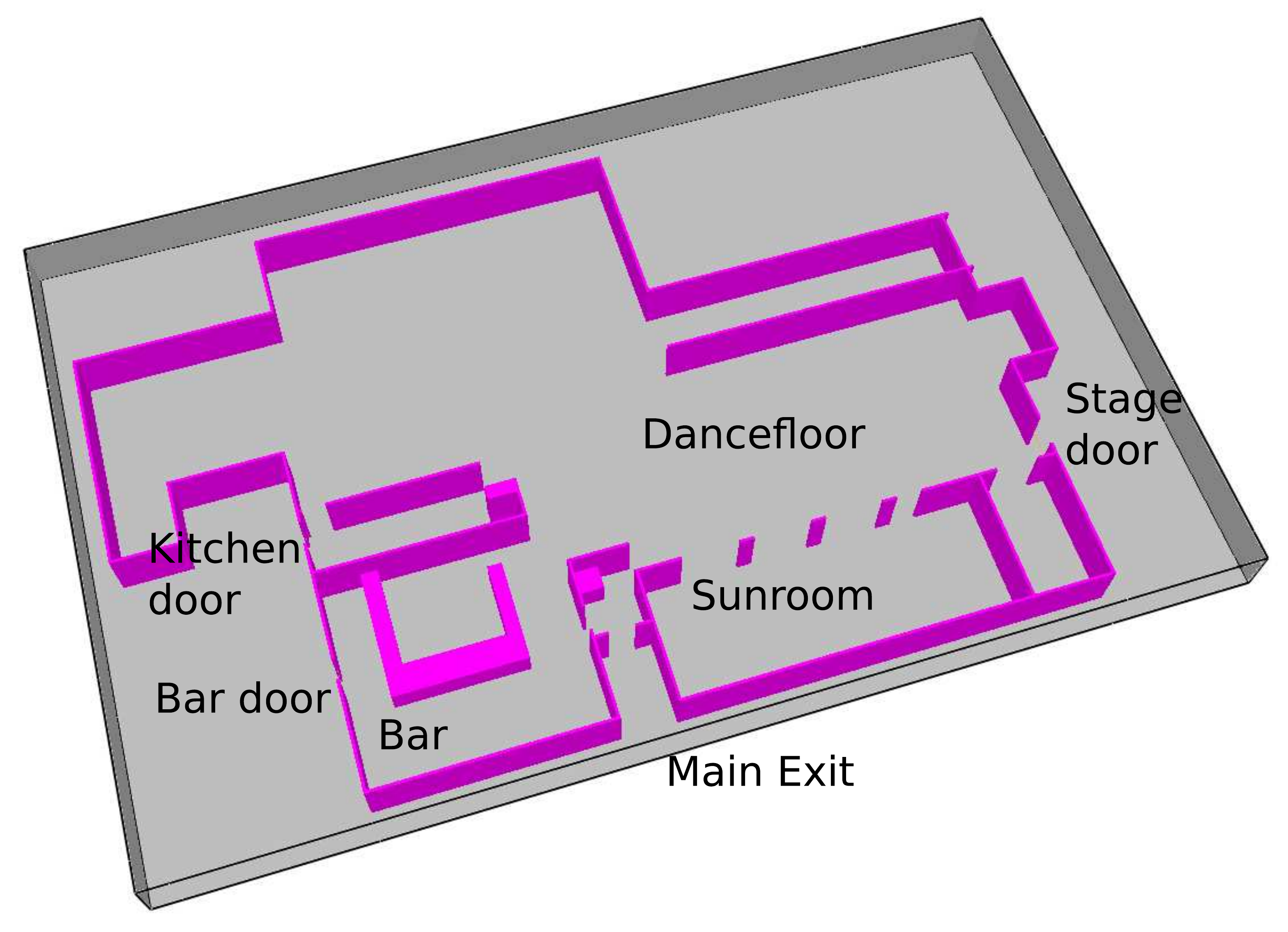}
		\caption{(Top) Floorplan of Station nightclub, taken from official report. (Bottom) Rendering in FDS+Evac.}
		\label{fig:render}
\end{figure}

We begin by rendering the  floor plan of the Station in FDS, using official architectural plans taken from \cite{station} (Figure~\ref{fig:render}). We use a figure of 450 for the number of agents to be simulated, and their initial distribution is specified according to \cite{station} (i.e., with high crowd densities in the Dancefloor and Sunroom areas, and lower densities in other areas).

We run two sets of experiments; the first, {\it idealised} set is designed to provide baseline evacuation data, and the second set replicates, as closely as possible, the conditions and events in the nightclub during the event. Investigation findings into the spread of the fire suggest that the Stage door became impassable 30 seconds from the start of the incident, so we reflect this fact in our simulation by closing that exit after that period has elapsed.  The official investigation was able to identify the exit paths for 248 of the 350 people who escaped from the building. The distribution of evacuees through the three other available exit routes was found to be non-uniform, with estimates of between one-half and two-thirds of patrons attempting to leave via the familiar main exit, rather than the under-utilised (and less familiar) main bar and Kitchen doors.  Reports suggest that only 12 people left via the Kitchen door during the evacuation.  In order to simulate this distribution of path choices, patrons are assigned a {\it probability of knowledge} for each exit route.  Exactly 12 evacuees are made aware of the existence of the Kitchen exit, and of the remaining patrons, 100\% are given knowledge of the main door, 50\% are given knowledge of the main bar door, and 25\% are given knowledge of the stage door. On the other hand, the idealised evacuation was structured as follows: there was no blocking of the Stage door, and agents in the simulation had full knowledge of all exit routes. This scenario represents the minimum time it would take to evacuate 450 people from the Station Nightclub, with optimum use made of available exit structures and no hindrance from fire, smoke, or unfavourable environmental conditions. 

We compare our simulation results with those obtained by the National Institute of Standards and Technology (NIST), and detailed in the official investigation report \cite{station}. In these experiments, NIST investigators used both Simulex \cite{thompson1995testing} and buildingEXODUS \cite{gwynne2001modelling} to evaluate both idealised and realistic evacuation scenarios. The results obtained were very similar for both packages, so we concentrate on the buildingEXODUS output. Within the ``realistic" simulation, occupants were instructed to always select the nearest exit, and the Stage door was also closed after 30 seconds. In the NIST simulation, 91 simulated occupants left via the building front door, which is precisely the number reported in the official investigation. Thirty-five simulated occupants used either the platform door or the kitchen door, which, again, is consistent with the evidence. 

\begin{figure}
		\centering
			\includegraphics[width=0.7\linewidth]{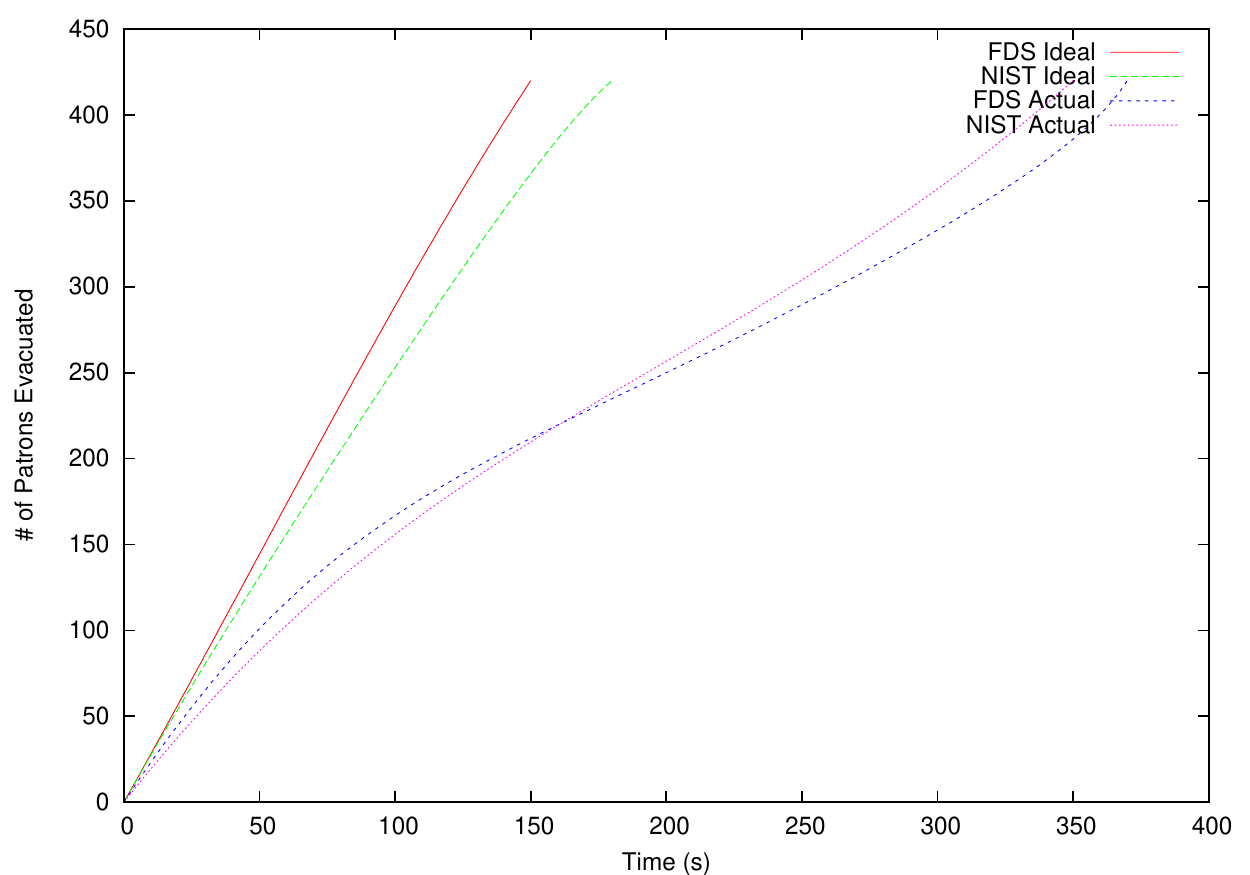}
		\caption{Comparison of leaving profiles between our simulation (FDS) and official NIST results.}
		\label{fig:valid}
\end{figure}

We therefore conclude that the official NIST simulations provide a sound basis for validating our own simulations. The results of the comparison are depicted in Figure ~\ref{fig:valid}. We note only that the results obtained (in terms of leaving profiles over time) are very similar to those reported by NIST, which supports the argument in favour of the soundness of our model.

\newpage
\subsection{Detection of crush}

Having established the validity of our simulation in terms of broad outcomes, the next stage is to specifically investigate the {\it emergence of crush}, and to see if this is easily detectable using Mutual Information. In order to achieve this, we measure the average force and the level of MI within our simulated population of 450 individuals, for both ``real" and ``idealised" evacuations. 

\begin{figure}
		\centering
			\includegraphics[width=0.7\linewidth]{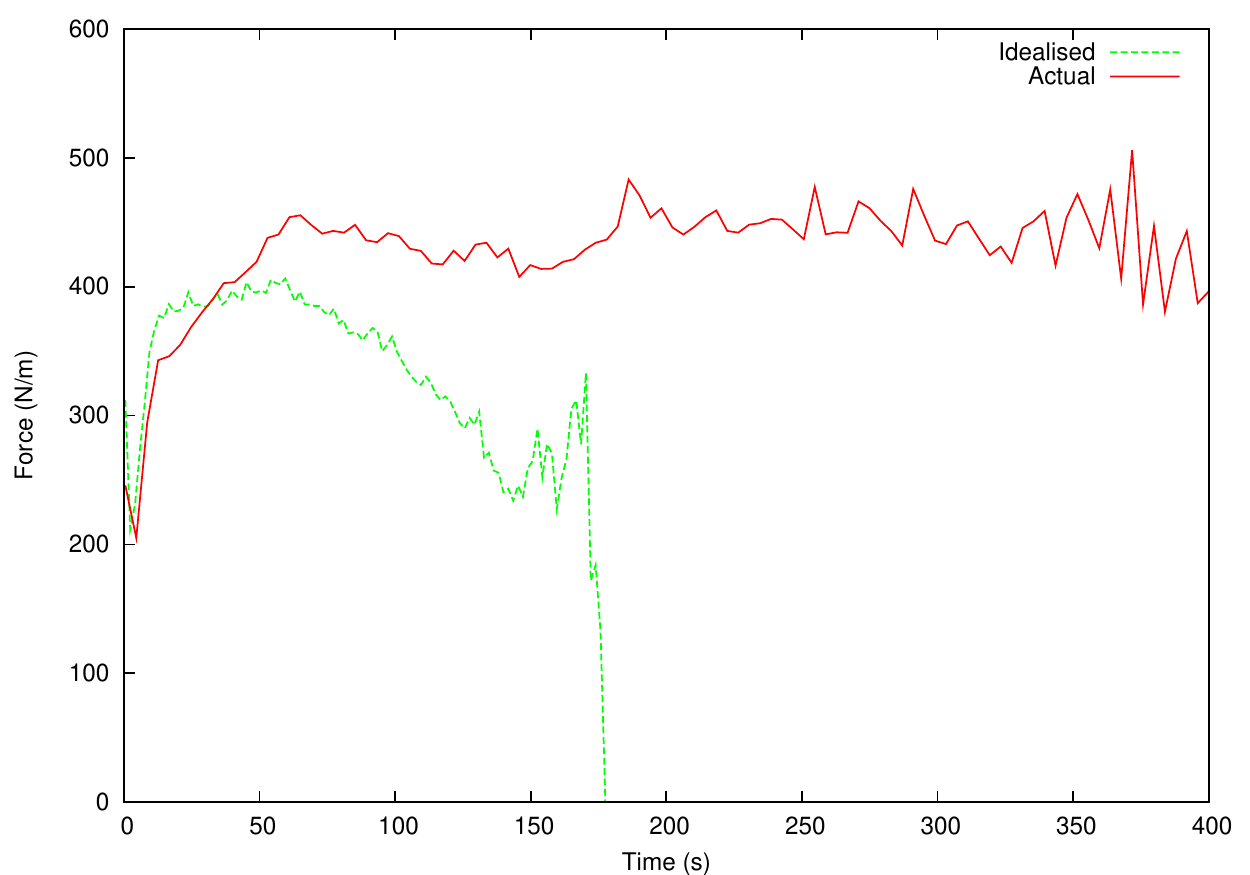}
		\caption{Comparison of average force between real and idealised scenarios.}
		\label{fig:force}
\end{figure}

We first consider the results of the force measurements, comparing them with evidence from the investigation. The force measurements for both scenarios are depicted in Figure ~\ref{fig:force}. Across both scenarios the levels of force initially increase as the evacuation commences, but it rapidly decays during the idealised version of events, since evacuees are more uniformly distributed. Force  levels drop to zero at around 175s, when everyone has left the building, which is broadly in line with the findings of the NIST idealised situation simulation (195s $\pm$ 7s).

In the ``real" scenario, we observe a sharp initial rise in average force, which initially peaks after around 65 seconds. This is directly in line with the findings of the official investigation, which states that a significant crowd crush occurred by the main entrance (where around a third of the fatalities occurred) at the beginning of the time period 71-102 seconds into the fire. 

\begin{quote}
Prior to 1-1/2 minutes into the fire, a crowd-crush occurred in the front vestibule which almost entirely disrupted the flow through the main exit. Many people became stuck in the prone position in the exterior double doors \cite[p. xx]{station}.

The camera angle shifts away from this door after 0:07:33 (0:01:11 fire time) and does not return to the front door until 0:08:04 (0:01:42 fire time). When the camera returns at 0:08:04 (0:01:42 fire time) a pile-up of occupants is visible. Details regarding how the pile-up occurred are not available from the WPRI-TV video; however, the interruption in flow of evacuating occupants apparent [in Figure 6-3] supports the contention that the disruption may have initiated early during the 31 second period when the camera was pointed elsewhere. \cite[p. 182]{station}
\end{quote}

\begin{figure}
		\centering
			\includegraphics[width=0.8\linewidth]{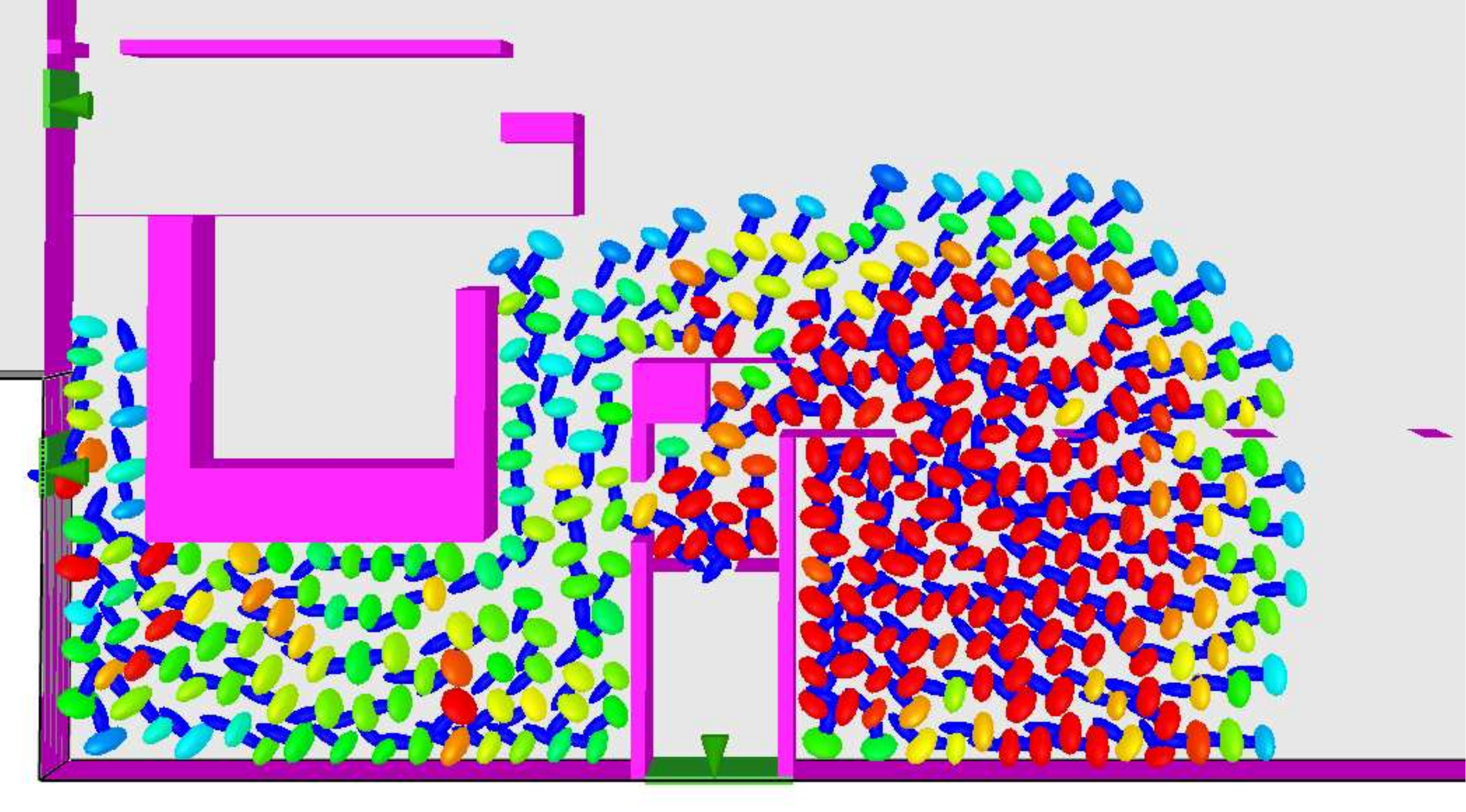}
		\caption{Screenshot of our fire scenario simulation after 65 elapsed seconds.}
		\label{fig:65s}
\end{figure}

In Figure ~\ref{fig:65s}, we show a screenshot of the simulation after 65 seconds, which graphically illustrates the significant crush around the main entrance and sunroom area (high levels of force are shown in red).

The analysis of MI during evacuation is performed using only {\it observable} variables, i.e. those with values that could be obtained via direct observation of an evacuation.  This is to ensure that our results were not \textit{implementation specific},  and to maximise the possibility of applying the technique in future  to other environments or video-captured data from \textit{real-life} evacuations.  Therefore, the three variables considered for analysis are the 2-dimensional {\it Cartesian coordinates} ($x_i$ and $y_i$) of each individual, $i$, together with their {\it heading} ($\Theta_i$).  We forego the use of {\it speed} within our analysis, as there is often little variation in speed during incidents with high population density.

We measure MI using Equation~\ref{eq:wicksmi}, taken from \cite{wicks2006a}:
	
	\begin{align}
		I(X,\Theta) &= \sum_{i,j} p(x_i,\theta_j) \log_2 \frac{p(x_i,\theta_j)}{p(x_i)p(\theta_j)} \nonumber \\
		I(Y,\Theta) &= \sum_{i,j} p(y_i,\theta_j) \log_2 \frac{p(y_i,\theta_j)}{p(y_i)p(\theta_j)} \nonumber \\
		I &= \frac{I(X,\Theta) + I(Y,\Theta)}{2}
	\label{eq:wicksmi}
	\end{align}

\begin{figure}
		\centering
			\includegraphics[width=0.7\linewidth]{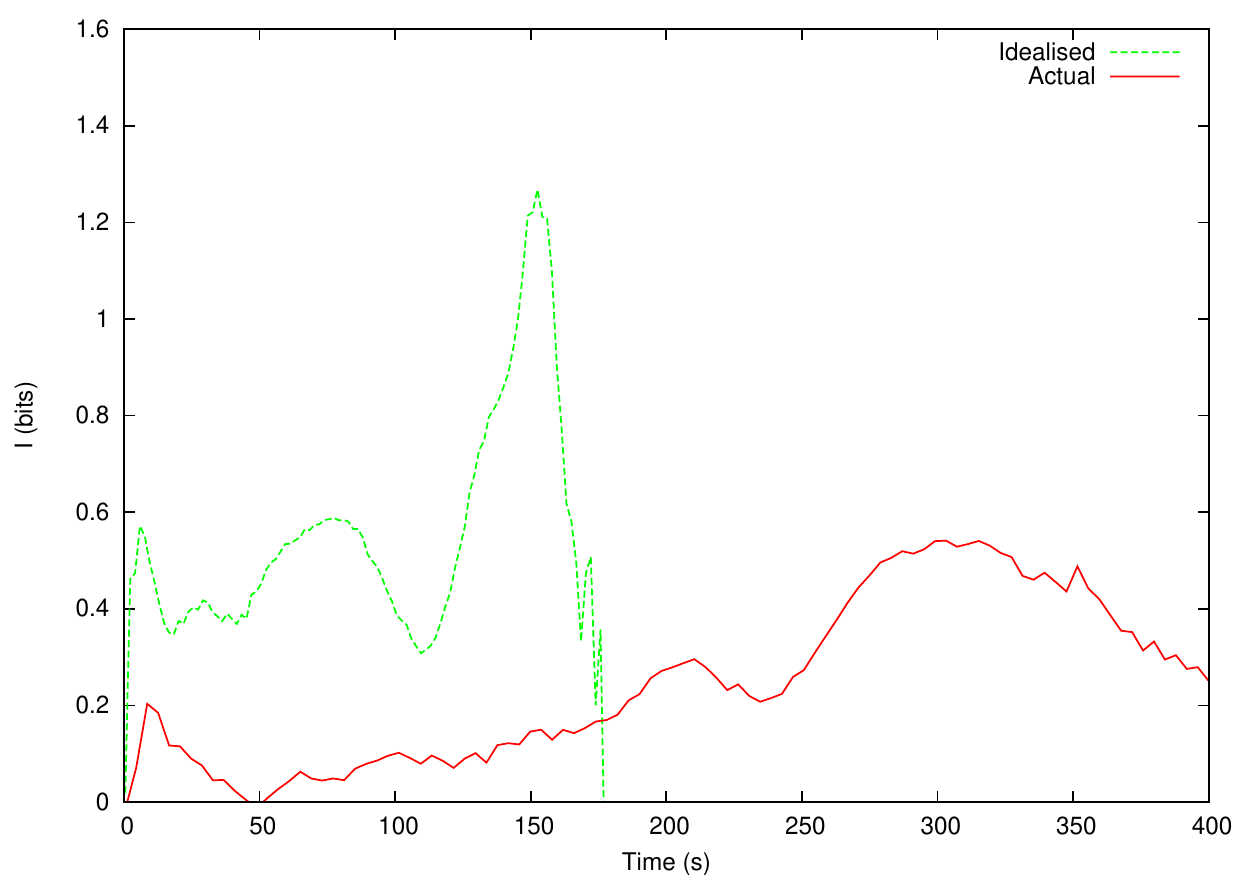}
		\caption{Comparison of Mutual Information between idealised and actual scenarios.}
		\label{fig:mi}
\end{figure}

Our MI measurements are depicted in Figure ~\ref{fig:mi}.
We expect to see, as the simulations begin, an initial rise in the MI of the system. As evacuees prepare to exit the structure they tend towards {\it alignment}, exhibiting similar escape trajectories to other evacuees in their locale.  In a maximally efficient evacuation this period of \textit{high order} (and high MI) would be sustained throughout, as evacuees would not alter their course in order to increase their chances of effective egress.  However, in an evacuation with a great deal of competition, the order in the system quickly breaks down, as the evacuees reposition themselves in order to increase their probability of escape.  MI may therefore may be used as an {\it order parameter}, where falling values of MI signify the breakdown of order within a specific evacuation. We observe marked quantitative differences in the MI readings between the two simulations.  During periods of disorder, MI should tend towards zero, whereas, during ordered segments of the evacuation, MI will rise significantly. 

\subsection{Idealised scenario}

In the idealised simulation, we see a sharp initial peak, as individuals all make for the exits at the same time. We then observe a drop, as the evacuees begin to compete for the available exit capacity.  An increase in order is seen as one exit route begins to clear, creating the rise in MI at $50 < t < 75$, falling back into a state of disorder as the final evacuees clear this (main bar) exit .  The MI reading then shows a progressive rise as the final evacuees exit the structure.  The sharp drop in MI at the end of the simulation occurs when the number of remaining evacuees falls below some (very low) threshold.

\newpage
\subsection{Realistic scenario}

The MI readings obtained from the simulation of \textit{actual} events show a far more disordered evacuation, with an initial rise in MI (signifying order) quickly disintegrating into {\it disorder}. The MI reading at $t \approx 50s$ approaches zero; this period of highly disordered evacuation remains as the exits to the structure are overwhelmed (see Figure~\ref{fig:65s}).  The exit rate of evacuees during this period is extremely low, which is confirmed by the exit profiles (see Figure~\ref{fig:valid}).  The MI level slowly rises towards the end of the evacuation, but, notably, the higher levels of order seen in the idealised evacuation are not reached until $t \approx 300s$, 5 minutes after the start of the evacuation.

\begin{figure}
	\centering
		\includegraphics[width=0.7\linewidth]{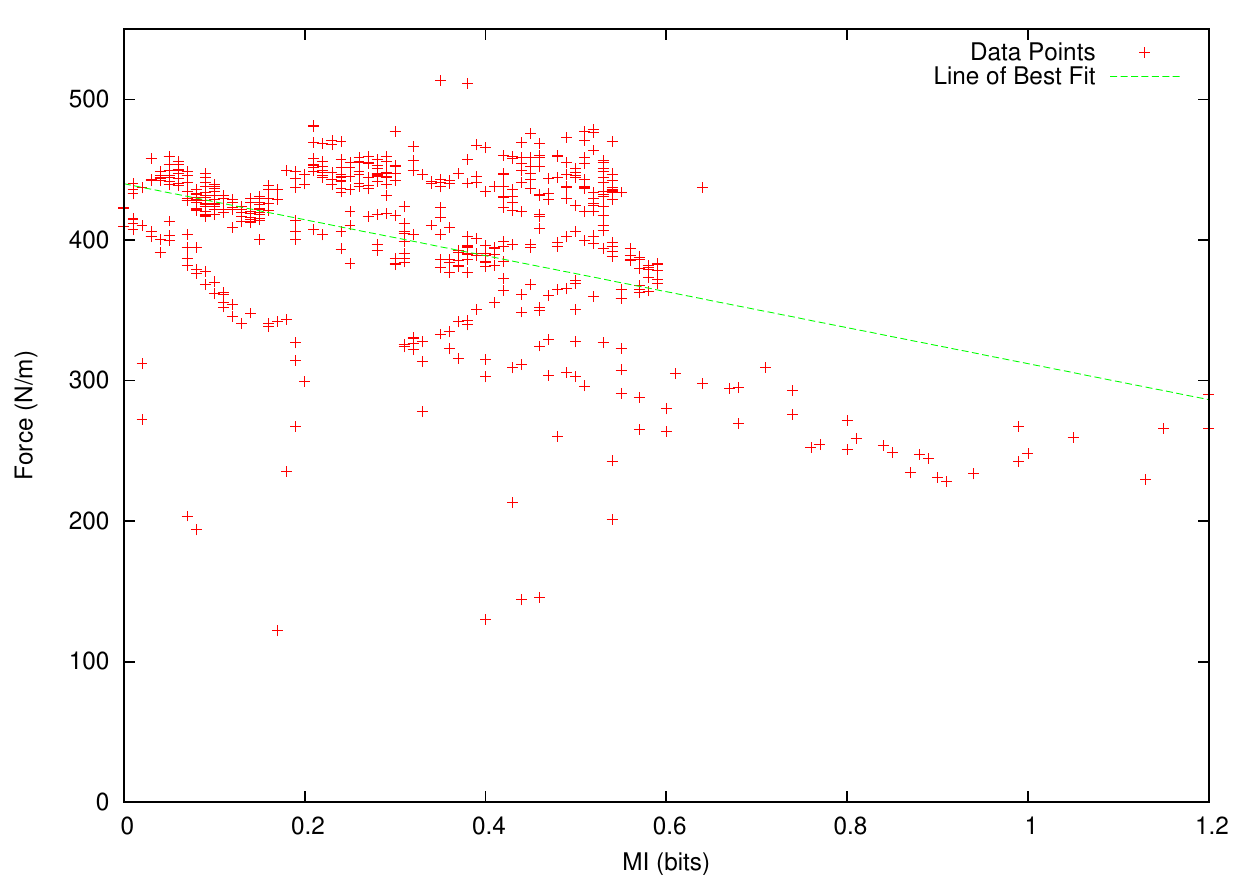}
		\caption{Scatterplot of Force versus Mutual Information.}
	\label{fig:polylobf}
	\end{figure}

\subsection{Correlation analysis}

We then perform a correlation analysis in order to establish the relationship (if any) between force and Mutual Information. A scatterplot of force versus MI suggests the existence of a statistical association (Figure ~\ref{fig:polylobf}), so we perform a simple linear correlation test.
The results of this are as follows:

	\begin{eqnarray*}
		P & = & 2.2e^{-16} \\
		R_p & = & -0.571 
	\end{eqnarray*}
	
The P-value obtained is much lower than the standard significance level for a two tailed test ($\alpha = 0.01$), ($P \ll \alpha$), which confirms the significance of the result. The correlation coefficient,  $R_p = -0.571$, confirms that there exists a negative correlation between MI and force within an evacuation scenario.

\section{Discussion and Conclusions}
\label{conc}

In this paper we have described a novel technique for the analysis of crowd evacuation scenarios. By calculating the Mutual Information of a system of interacting individuals, we are able to determine the level of internal force present within a crowd. We have shown that consistently low levels of Mutual Information are correlated with high levels of force within a crowd. This method removes the need for computationally expensive physical force calculations, and allows planners to quickly and easily incorporate objective measures of crowd disorder and crush into their simulation scenarios. Future work will focus on refinements of the technique, as well as investigation of its ``real-world" applicability. We are particularly interested in the potential for using our technique to analyse real-time video images, with the eventual aim of developing an on-site automatic early warning system for crush and disorder at large-scale events.

\bibliographystyle{plain}
\bibliography{micrush}
\end{document}